\documentclass[12pt]{article}
\usepackage{latexsym}
\begin{document}

\newcommand{\be}{\begin{equation}}
\newcommand{\ee}{\end{equation}}
\newcommand{\bea}{\begin{eqnarray}}
\newcommand{\eea}{\end{eqnarray}}
\newcommand{\gt}{\widetilde{g}}
\newcommand{\Tt}{\widetilde{T}}
\newcommand{\lag}{{\cal L}}
\newcommand{\rhot}{\widetilde{\rho}}
\newcommand{\pt}{\widetilde{p}}
\newcommand{\rhodyn}{\rho_{\rm dyn}}
\newcommand{\pdyn}{p_{\rm dyn}}
\newcommand{\4}{{(4)}}
\newcommand{\5}{{(5)}}
\newcommand{\M}{{\rm M}}
\newcommand{\R}{{\rm R}}

\baselineskip 18pt

\begin{flushright}
hep-th/0105007 \\ EFI-2000-53  \end{flushright} 

\vspace*{2cm}

\begin{center}

{\Large{\bf Can we live in a self-tuning universe?}}

\vspace*{0.3in}
Sean M. Carroll$^{1}$ and Laura Mersini$^{2}$
\vspace*{0.3in}

\it
$^{1}$Enrico Fermi Institute and Department of Physics,
University of Chicago\\
5640 S.~Ellis Avenue, Chicago, IL~60637, USA\\
{\tt carroll@theory.uchicago.edu} \\
\vspace*{0.2in}

$^{2}$Scuola Normale Superiore, Piazza dei Cavalieri 7, I-56126
Pisa, Italy \\
{\tt mersini@cibs.sns.it}\\
\vspace*{0.2in}

\end{center}

\begin{abstract}
The self-tuning brane scenario is an attempt to solve the cosmological
constant problem in the context of extra dimensions.  Rather than
making the vacuum energy small, this approach proceeds by removing the
gravitational effect of vacuum energy on the expansion of the
universe.  Such behavior is only possible through changing the
Friedmann equation of conventional cosmology, and we discuss
difficulties in obtaining cosmological evolution compatible with
observation in this context.  Specific models considered include a
bulk scalar field coupling to the brane via a conformal transformation
of the brane metric, and via a rescaling of the brane volume element.
\end{abstract}

\vfill

\newpage

\section{Introduction}

The fact that the observed cosmological constant is much smaller
than the expected value \cite{weinberg,witten,carroll} may provide
a crucial clue in our attempts to understand the nature of spacetime.
Most attempts to solve the problem can be characterized
as making the vacuum energy much smaller than its natural value.
Alternatively, however, we can imagine keeping a large vacuum
energy, but changing the gravitational dynamics in such a way that
the vacuum does not act as a (significant) source of spacetime curvature.
Since it is only through its gravitational influence that
the vacuum energy can be measured, 
such an arrangement could reconcile the
naive estimates $\rho_{\rm vac}\geq (10^{18}~{\rm GeV})^4$
with the observationally favored result
$\rho_{\rm vac} \sim (10^{-3}~{\rm eV})^4$ \cite{sne,carroll}.

The idea of brane-worlds and large extra dimensions
\cite{extra} opens up
a new set of ways to think about the cosmological constant
problem.  In these scenarios, our observed four-dimensional theory
of gravity is descended from a higher-dimensional embedding, and
in principle the resulting dynamics can differ dramatically from
a straightforward four-dimensional expectation. 

An example of such an altered dynamics is provided by the idea of
self-tuning branes 
\cite{Arkani-Hamed:2000eg,Kachru:2000}.
Here, matter fields on a three-brane with a single extra dimension 
are coupled to a bulk scalar field.  
With an appropriate choice of couplings, Minkowskian
solutions on the brane can be found with any brane cosmological
constant.  In this paper, we put aside fundamental issues of
the feasibility of the self-tuning idea ({\it e.g.}, the role
of singularities) to concentrate on whether this kind
of scenario can be made compatible with conventional cosmology.
(See \cite{Binetruy:2000ut,bc} for investigations of brane-world cosmology, 
and \cite{stc,Mennim:2000wv} for studies of the self-tuning scenario.)

On a self-tuning brane, there is vacuum energy (essentially the
tension of the brane), but the spacetime geometry is nevertheless
flat.  Somehow, then, the geometry is insensitive to the vacuum
energy.  It is necessary, however, that spacetime respond to at least
some sorts of energy density; in particular, the success of 
Big Bang Nucleosynthesis (BBN) \cite{bbn} provides evidence in
favor of the standard picture in the radiation-dominated era.
It is therefore necessary to recover at least some portion of
conventional cosmology, while removing the effects of the 
cosmological constant.

In this work we attempt to characterize the empirical challenges
to a successful self-tuning cosmology.  In the next section we
discuss in general how the self-tuning mechanism may be 
understood in terms of the energy and pressure on the brane,
pointing out the distinctions with conventional Friedmann
cosmology.  We then 
consider two specific models of self-tuning, and derive effective
Friedmann-like equations relating the Hubble parameter to the
energy and pressure.  In section \ref{discussion} we compare
these models with what we know about the universe, and discuss 
whether they may be brought into agreement with observation.  
Although such agreement seems unlikely, we are unable to rule
it out entirely.

\section{The secret of self-tuning cosmology}
\label{abstract}

In this section we consider how a theory of gravity may in
principle be insensitive to vacuum energy while allowing other
forms of energy-momentum to influence spacetime curvature.
We consider a flat Robertson-Walker metric in $3+1$ dimensions,
\be
  ds^2 = -dt^2 + a^2(t)d{\vec x}^2 \ ,
  \label{rwmetric}
\ee
where $a(t)$ is the scale factor and $d{\vec x}^2$ is the 
flat Euclidean metric.  An energy-momentum tensor
consistent with a Robertson-Walker metric will be spatially isotropic, 
taking the form
\be
  T^\mu{}_\nu = {\rm diag}(-\rho, p, p, p)\ ,
  \label{tmunu}
\ee
where $\rho$ is the energy density and $p$ the pressure.
The conventional Friedmann equation of general relativity is
then
\be
  H^2 \equiv \left({\dot a\over a}\right)^2 = {8\pi G\over 3}\rho\ ,
  \label{feq}
\ee
where $H$ is the Hubble parameter and $G$ is the (four-dimensional)
Newton's constant.

In a self-tuning model, it is possible to find a Minkowski-spacetime
solution ($H=0$) regardless of the value of $\rho_{\rm vac}$ (when
all other energy densities vanish).  Clearly, for this to be 
possible (\ref{feq}) will have to be modified, either in its
explicit form or in the definition of $\rho$.  If, however, we
consider theories which arise from varying a specified action
with respect to the metric tensor to derive gravitational field
equations, the energy density and pressure (in this coordinate
system) are defined by
\be
  \rho = -T^0{}_0 = 2{1\over \sqrt{g}}g^{00}{\partial\lag_{\rm matter}
  \over \partial g^{00}}\ ,\qquad
  p = T^1{}_1 = -2{1\over \sqrt{g}}g^{11}{\partial\lag_{\rm matter} \over 
  \partial g^{11}}\ ,
\ee
where $g$ is the absolute value of the determinant of the metric and 
$\lag_{\rm matter}$ is the matter Lagrange density.  These definitions
do not discriminate between different forms of energy density;
there is no way for the gravitational field to tell the difference
between energy density from the vacuum and energy density from any
other source.

How, then, can we modify the theory to allow flat solutions in
the presence of vacuum energy, while remaining sensitive to the
influence of other sources on the expansion rate?  
Given the metric (\ref{rwmetric})
and energy-momentum tensor (\ref{tmunu}), we seek an equation 
written in terms of $\rho$, $p$, $a$ and its derivatives (which
appear in the curvature tensor), and possibly explicit additional
fields.  Although vacuum energy should enter any such equation
in the same way as other energy, the vacuum does have a 
distinguishing characteristic, namely its equation of state:
\be
  p_{\rm vac} = -\rho_{\rm vac}\ .
\ee
With the ingredients at our disposal, this relation suggests
a form for a modified Friedmann equation:
\be
  H^2 = f(\rho, p) (\rho + p) + {\rm other~terms}\ ,
  \label{modf}
\ee
where $f(\rho, p)$ is a well-behaved function at $p=-\rho$.  Such
a relation would allow for a Minkowski solution ($H=0$) in the
presence of arbitrary vacuum energy.

Although (\ref{modf}) differs from the conventional Friedmann law,
it is not necessarily incompatible with observation.  Consider for
example a hypothetical relation
\be
  H^2 = 2\pi G (\rho + p)\ .
  \label{wish}
\ee
During a matter-dominated era, $p=0$, and the Hubble parameter will
differ by a factor $\sqrt{3}/2\approx 0.87$ for a given value of the 
energy density.  However, since the scaling of $H$ with $\rho$ is as in
the conventional theory, and $\sqrt{3}/2$ is close to unity, it 
would be hard to distinguish between the behavior predicted by 
(\ref{wish}) and the usual Friedmann equation during matter
domination.  During radiation domination, $p=\rho/3$, and 
(\ref{wish}) precisely recovers the conventional expectation
(by construction).  This is important, as our most precise quantitative
evidence in favor of conventional cosmology comes from BBN,
which occurs while the universe is 
radiation-dominated.\footnote{The anisotropy spectrum in the
cosmic microwave background (CMB), although it does provide precision
constraints on the expansion of the universe, does not do so in a
model-independent way.  Since a novel theory of gravity which 
predicted a different Friedmann equation could also predict 
different behavior for the evolution of large-scale density
perturbations, it is impossible to compare directly a phenomenological
relationship such as (\ref{wish}) to CMB observations.  In any specific
full theory, CMB anisotropies are likely to provide an interesting
test.}  Finally,
this relation would solve the cosmological constant problem, as
$H=0$ when $p=-\rho$. 

Unfortunately, we do not have a theory that predicts (\ref{wish}).
As we shall see below, the specific self-tuning scenarios we consider
lead to additional derivatives of the metric, as well as explicit
dependence on the bulk scalar field, and extra terms which are 
quadratic in the energy and pressure.  Generally, it seems unlikely
that a theory which did predict a linear dependence of $H^2$ on
$(\rho+p)$ would both get the correct coefficient $2\pi G$ (or
very close to it) and successfully recover Newton's law
$F = Gm_1m_2/r^2$ in the solar system.  However, this unlikelihood
does not seem so great that searching for such a theory would be
a waste of time.

\section{Specific Examples}

%%%%%%%%%%%%%%%%%%%%%%%%%%%%%%%%%%%%%%%%%%%%%%%%%%%%%%%%%%%%%%%%%%%%%%%%%
\subsection{Setup}
\label{setup}

We start with an action of the form \cite{Arkani-Hamed:2000eg,Kachru:2000}
\be
  S = S_5[\phi,g^\5] + S_4[\phi, g^\4, \psi_i]\ .
  \label{action}
\ee
Here, $S_5$ is the bulk action and $S_4$ that of the brane; 
$\phi$ is a scalar field in the bulk with interactions on the
brane, while the $\psi_i$'s are matter fields confined to the
brane.  The metric in the bulk is $g^\5_{ab}$, and 
we choose coordinates such that the induced metric on
the brane is
\be
  g^\4_{\mu\nu} = \delta_\mu^a \delta_\nu ^b g_{ab}\ .
\ee
The indices
$a,b$ run over $\{0,1,2,3,y\}$ and $\mu,\nu$ run over $\{0,1,2,3\}$.

The bulk action can be written
\be
  S_5 = \int d^5x\, \lag_5 = \int d^5x\, \sqrt{g^\5}
  \left({M_5^3\over 2}R -
  \alpha(\nabla\phi)^2\right)\ ,
  \label{action5}
\ee
with $\alpha$ a coefficient which we leave unspecified for
the moment, $M_5$ is the five-dimensional Planck mass, $R$
is the five-dimensional Ricci scalar, and the brane action as
\be
  S_4 = \int d^5x\, \lag_4 = \int d^5x\, \sqrt{g^\4}
  \widehat{\lag_4} \delta(y) = \int d^4x\, \sqrt{g^\4}
  \widehat{\lag_4} \ .
  \label{action4}
\ee
We will leave the actual form of $\widehat{\lag_4}(\phi, g^\4, \psi_i)$
unspecified in this section, and consider different possibilities in
sections \ref{conformal} and \ref{volume}.  We will always consider
geometries that have a ${\bf Z}_2$ symmetry with the brane at the fixed
point, so that we need not include a boundary
term in the brane action.

The 5-dimensional Einstein's equations are 
\be
  G_{ab} = M_5^{-3} T_{ab}\ ,
\ee
where the energy-momentum tensor is defined by
\be
  T_{ab} = -2 {1\over \sqrt{g^\5}}{\partial \lag \over \partial g_\5^{ab}}\ .
\ee
We choose the metric
\be
  ds_\5^2 = -n^2(y,t)dt^2 + a^2(y,t)d{\vec x}^2 + b^2(y,t)dy^2\ ,
\ee
where ${\vec x} = \{x^1, x^2, x^3\}$ are the spatial coordinates
along the brane, and $y$ is the transverse spatial dimension.
The Einstein tensor is then
\begin{eqnarray}
  G_{00} &=& 3\left[{{\dot a}^2 \over a^2}
  + {{\dot a}\dot b \over ab} - {n^2 \over b^2} 
  \left({a'' \over a} + {{a'}^2 \over a^2} 
  - {a' b' \over ab}\right)\right], 
  \label{G00} \\
  G_{0i}  &=& 0
  \\
  G_{0y} &=&  3\left(- {\dot{a}^{\prime} \over a}
  + {a' \dot b \over ab} + {{\dot a} n' \over an} 
  \right),
  \label{G0y} \\
  G_{ij} &=&
  {a^2 \over n^2} \left(- 2{\ddot a \over a} - {\ddot b \over b} 
  -{{\dot a}^2 \over a^2} - 2{{\dot a}\dot b \over ab}
  + 2{{\dot a}\dot n \over an}
  + {\dot b \dot n \over bn} \right)\delta_{ij}
  \nonumber \\
  & & + {a^2 \over b^2} \left( 2 {a'' \over a} + {n'' \over n} 
  + {{a'}^2 \over a^2} - 2{a'b' \over ab} + 2{a'n' \over an} 
  - {b' n' \over bn} 
  \right)
   \delta_{ij}\ ,  
  \label{Gij} \\
  G_{iy}  &=& 0
  \\
  G_{yy} &=& 3\left[ {b^2 \over n^2} 
  \left(-{\ddot a \over a} - {{\dot a}^2 \over a^2} 
  + {\dot a \dot n \over an} \right) 
  + {{a'}^2 \over a^2} + {a' n' \over an} \right].
  \label{Gyy} 
\end{eqnarray} 

The energy-momentum tensor decomposes into contributions from the
bulk and the brane,
\be
  T_{ab} = T^{(5)}_{ab} + T^{(4)}_{ab} = 
  -2 {1\over \sqrt{g^\5}}\left({\partial \lag_5 \over \partial g_\5^{ab}}
  + {\partial \lag_4 \over \partial g_\4^{\mu\nu}}
  \delta^\mu_a \delta^\nu_b\right)  \ .
  \label{tab}
\ee
The components of the bulk $T^{(5)}_{ab}$ receive contributions
only from the scalar field $\phi$, given by
\begin{eqnarray}
  T^{(5)}_{00} &=& \alpha n^2 \left({{\dot\phi}^2 \over n^2} 
  + {{\phi^\prime}^2 \over b^2}\right)
  \\
  T^{(5)}_{0y} &=& 2\alpha \dot\phi \phi^\prime
  \\
  T^{(5)}_{ij} &=& \alpha a^2 \left({{\dot\phi}^2 \over n^2} 
  - {{\phi^\prime}^2 \over b^2}\right)\delta_{ij}
  \\
  T^{(5)}_{yy} &=& \alpha b^2 \left({{\dot\phi}^2 \over n^2} 
  + {{\phi^\prime}^2 \over b^2}\right)\ .
\end{eqnarray}
The contribution to the energy-momentum tensor from the brane will
depend on the chosen brane action; in terms of $\widehat{\lag_4}$
defined in (\ref{action4}), we have
\be
  T^{(4)}_{ab} = \sqrt{g^\4 \over g^\5}
  \left(g^\4_{\mu\nu}\widehat{\lag_4} -2 {\partial \widehat{\lag_4} 
  \over \partial g_\4^{\mu\nu}}\right) \delta(y) \delta^\mu_a \delta^\nu_b
  \ . \label{tab4}
\ee

In addition to Einstein's equations,
we need the equation of motion for $\phi$.  With the action
(\ref{action5})-(\ref{action4}), the $\phi$ equation is
\be
  \nabla_a {\partial {\cal L}_5 \over \partial
  (\nabla_a \phi)} =
  {\partial {\cal L}_4 \over \partial \phi}\ ,
  \label{phi1}
\ee
since we will be considering brane actions that depend on $\phi$ but not its
derivatives.
The left-hand side is
\be
  \nabla_a {\partial {\cal L}_5 \over \partial
  (\nabla_a \phi)} = -2\alpha\sqrt{g_5}\Box\phi\ ,
\ee
yielding an equation of motion
\be
  \Box\phi = -{1\over 2\alpha} {1\over b} %\sqrt{g^\4 \over g^\5}
  {\partial \widehat{\cal L}_4 \over \partial \phi} \delta(y)\ ,
  \label{phi2}
\ee
where $\widehat{\cal L}_4$ is defined by (\ref{action4})
and the D'Alembertian on scalars is given by
\be
  \Box\phi = -{1\over n^2}\left[ \ddot\phi + \left( -{\dot n \over n}
  + 3 {\dot a \over a} + {\dot b \over b}\right)\dot\phi\right]
  + {1\over b^2}\left[ \phi'' + \left( {n' \over n}
  + 3 {a' \over a} - {b' \over b}\right)\phi'\right]\ .
  \label{box}
\ee

The equations of motion generally involve distributional sources
localized on the brane.  It is therefore convenient to separate the 
equations into distinct relations valid in the bulk and on the brane.
The bulk equations are simply the full equations with distributional
sources omitted; they include the bulk Einstein equations:
\begin{eqnarray}
  3\left[{{\dot a}^2 \over a^2}
  + {{\dot a}\dot b \over ab} - {n^2 \over b^2} 
  \left({a'' \over a} + {{a'}^2 \over a^2} 
  - {a' b' \over ab}\right)\right] &=&
  {\alpha\over M_5^3} n^2 \left({{\dot\phi}^2 \over n^2} 
  + {{\phi^\prime}^2 \over b^2}\right)
  \label{00} \\
  3\left(- {\dot{a}^{\prime} \over a}
  + {a' \dot b \over ab} + {{\dot a} n' \over an} 
  \right)  &=&
  {2\alpha\over M_5^3} \dot\phi \phi^\prime 
  \label{0y} \\
  {a^2 \over n^2} \left(- 2{\ddot a \over a} - {\ddot b \over b} 
  -{{\dot a}^2 \over a^2} - 2{{\dot a}\dot b \over ab}
  + 2{{\dot a}\dot n \over an}
  + {\dot b \dot n \over bn} \right) \qquad\qquad
  \nonumber \\
  + {a^2 \over b^2} \left( 2 {a'' \over a} + {n'' \over n} 
  + {{a'}^2 \over a^2} - 2{a'b' \over ab} + 2{a'n' \over an} 
  - {b' n' \over bn} \right) 
  &=& {\alpha\over M_5^3} a^2 \left({{\dot\phi}^2 \over n^2} 
  - {{\phi^\prime}^2 \over b^2}\right)
  \label{ij} \\
  3\left[ {b^2 \over n^2} 
  \left(-{\ddot a \over a} - {{\dot a}^2 \over a^2} 
  + {\dot a \dot n \over an} \right) 
  + {{a'}^2 \over a^2} + {a' n' \over an} \right] &=&
  {\alpha\over M_5^3} b^2 \left({{\dot\phi}^2 \over n^2} 
  + {{\phi^\prime}^2 \over b^2}\right)\ ,
  \label{yy}
\end{eqnarray} 
and the bulk scalar field equation
\be
  -{1\over n^2}\left[ \ddot\phi + \left( -{\dot n \over n}
  + 3 {\dot a \over a} + {\dot b \over b}\right)\dot\phi\right]
  + {1\over b^2}\left[ \phi'' + \left( {n' \over n}
  + 3 {a' \over a} - {b' \over b}\right)\phi'\right]
  = 0\ .
  \label{phi}
\ee 
The corresponding equations on the brane can be expressed
without explicit $\delta$-functions by integrating them in the
vicinity of the brane to obtain jump conditions
\cite{Binetruy:2000ut}.
Assuming ${\bf Z}_2$ symmetry, the jump conditions
relate second derivatives with respect to $y$ to the
coefficients of $\delta$-function sources.  That is, expressions
of the form
\be
  f''(y) = A\delta(y) + B
\ee
imply, on the brane,
\be
  f'_0 = {1\over 2} A\ ,
  \label{jump}
\ee
where a subscript $0$ indicates that a quantity is evaluated
on the brane by taking the limit as $y\rightarrow 0^+$.  
For the scalar field, (\ref{phi2}) and (\ref{box})
imply
\be
  \phi'_0 = -{1\over 4\alpha} b_0 
  \left({\partial \widehat{\cal L}_4 \over \partial \phi}\right)_0\ .
  \label{phijump}
\ee
For the metric, it is convenient to express the jump conditions
for $a'_0$ and $n'_0$ separately for each model, which we do below.

To get a Friedmann-like equation on the brane, we consider
the $yy$ component of Einstein's equations (\ref{yy}), evaluated
on the brane.  (There is no $\delta$-function in $T_{yy}$, so the
bulk equation (\ref{yy}) holds true on the brane as well.)
We can scale our time coordinate such that
$n_0=1$, ${\dot n}_0 = 0$ (although $n$ will generally vary
off the brane).  We then have
\be
  {\ddot a_0 \over a_0} + {{\dot a_0}^2 \over a_0^2} =
  {{a_0'}^2 \over a_0^2b_0^2}
  + {a_0' n_0' \over a_0b_0^2}
  - {\alpha \over 3M_5^3} {{\phi_0^\prime}^2 \over b_0^2}
  - {\alpha \over 3M_5^3} {\dot\phi_0}^2\ .
  \label{f1}
\ee
The next step is to express the spatial derivatives on the right
hand side of (\ref{f1}) in terms of energy and momentum on the brane,
which requires a specification of the brane Lagrangian ${\cal L}_4$.
In the original papers on self-tuning branes \cite{Arkani-Hamed:2000eg,
Kachru:2000}, the matter Lagrangian was taken to be a cosmological
constant $\lambda$ times an exponential of $\phi$: 
\be
  {\cal L}_4 = -\sqrt{g^\4} e^{2\beta\phi}
  \lambda \delta(y)\ ,
  \label{originalL}
\ee
with $\beta$ a coupling constant.
In what follows we consider two distinct
generalizations of this form to actions with dynamical fields:
conformal coupling, in which matter fields on the brane couple exclusively
to a rescaled metric $\widetilde{g}_{\mu\nu} = e^{\beta \phi}g^\4_{\mu\nu}$,
or volume element rescaling, in which the four-dimensional volume element
is taken to be $d^4x\, \sqrt{g^\4} e^{2\beta\phi}$ (and the four-dimensional
metric is otherwise simply the induced metric from five dimensions).  Both 
approaches reduce to (\ref{originalL}) when brane tension is the only 
contribution to the four-dimensional action, but give different results for
other forms of matter.

%%%%%%%%%%%%%%%%%%%%%%%%%%%%%%%%%%%%%%%%%%%%%%%%%%%%%%%%%%%%%%%%%%%%%%%%%
\subsection{Conformal coupling}
\label{conformal}

In this section we consider an action of the form
\be
  S = S_5[\phi,g^\5_{ab}] + S_4[\psi_i, e^{\beta\phi}g^\4_{\mu\nu}]\ ,
\ee
where the five-dimensional action is as in (\ref{action5}), and the
four-dimensional action is
\be
  S_4 = \int d^5x\, \lag_4
  = \int d^5x\, \sqrt{\gt} f(\psi_i, \gt_{\mu\nu})\delta(y) 
  = \int d^4x\, \sqrt{\gt} f(\psi_i, \gt_{\mu\nu}) \ .
\ee
Here, $\psi_i$ represents an unspecified set of matter fields, and
the metric to which matter on the brane couples is related to
the induced metric by a conformal transformation,
\be
  \gt_{\mu\nu}= e^{\beta\phi}g^\4_{\mu\nu}\ ,
\ee
equivalent to $\widehat{\cal L}_4=e^{2\beta\phi}f(\psi_i, \gt_{\mu\nu})$
in the notation defined by (\ref{action4}).

In this model, test particles on the brane move along geodesics
of $\gt_{\mu\nu}$; this is the only metric perceived by observers
on the brane.  It therefore is sensible to define all brane
quantities in terms of this metric.
The energy-momentum tensor as measured by observers
living on the brane takes the form
\be
   \Tt_{\mu\nu} \delta(y) = -2 {1\over \sqrt{\gt}}
  {\partial \lag_4 \over \partial \gt^{\mu\nu}}\ .
\ee
The energy density and pressure as seen by brane observers
will be, in our coordinate system,
\be
  \rhot = -\gt^{00} \Tt_{00}
\ee
and
\be
  \pt = \gt^{11} \Tt_{11}\ ,
\ee
so we have
\begin{eqnarray}
  \Tt_{00} &=&  n^2 e^{\beta\phi}\rhot\\
  \Tt_{ij} &=&  a^2 e^{\beta\phi} \pt \delta_{ij} \ .
\end{eqnarray}

The energy-momentum tensor to which the five-dimensional Einstein's
equations couple is given by (\ref{tab}).
Since the brane action depends on $g_{\mu\nu}$ only through
$\gt_{\mu\nu} = e^{\beta\phi}g_{\mu\nu}$, we can write
\be
  {\partial \lag_4 \over \partial g^{\mu\nu}} = 
  {\partial \gt^{\rho\sigma} \over \partial g^{\mu\nu}}
  {\partial \lag_4 \over \partial\gt^{\rho\sigma}}
  = -{1\over 2} e^{-\beta\phi} \sqrt{\gt} \Tt_{\mu\nu} \delta(y)\ .
\ee
Using $\sqrt{g}=na^3b$ and $\sqrt{\gt}= e^{2\beta\phi}na^3$, 
from (\ref{tab}) we get
\begin{eqnarray}
  T^{(4)}_{ab} &=& e^{-\beta\phi} \sqrt{\gt \over g} 
  \Tt_{\mu\nu}\delta^\mu_a \delta^\nu_b \delta(y) \nonumber \\
  &=& {e^{\beta\phi}\over b} \Tt_{\mu\nu}
  \delta^\mu_a \delta^\nu_b \delta(y) \ ,
\end{eqnarray}
or, more explicitly,
\begin{eqnarray}
  T^{(4)}_{00} &=& {n^2 \over b} e^{2\beta\phi} \rhot \delta(y) 
  \label{t400} \\
  T^{(4)}_{ij} &=& {a^2 \over b} e^{2\beta\phi} \pt\delta(y)\delta_{ij}\ ,
  \label{t4ij}
\end{eqnarray}
with other components vanishing.  

Our primary interest is in the cosmological equation (\ref{f1}),
where we are now in a position to evaluate the first derivatives
of the metric coefficients $a$ and $n$ on the brane.  These are
derived using the general relation (\ref{jump}) applied to 
Einstein's equations $G_{ab} = M_5^{-3}T_{ab}$, with $G_{ab}$
given by (\ref{G00}-\ref{Gyy}) and the relevant components
of $T_{ab}$ by (\ref{t400}-\ref{t4ij}).
The jump conditions for the metric components then yield
\begin{eqnarray}
  {a_0' \over a_0} &=& -{1\over 6M_5^3} b_0 e^{2\beta\phi_0}\rhot\ , 
  \label{jump1}\\
  {n_0' \over n_0} &=& {1\over 6M_5^3} b_0 e^{2\beta\phi_0}
  (2\rhot + 3\pt)\ .
  \label{jump2}
\end{eqnarray} 

Finally we need the jump condition for $\phi$, derived from 
the equation of motion (\ref{phi2}).
In the case of conformal coupling the brane source for $\phi$
can be expressed in terms of $\rhot$ and $\pt$ by using
the fact that ${\cal L}_4$ depends on $\phi$
only through $\gt_{\mu\nu}$ to write
\begin{eqnarray}
  {\partial {\cal L}_4 \over \partial \phi} = 
  {\partial \gt^{\mu\nu} \over \partial \phi}
  {\partial {\cal L}_4 \over \partial \gt^{\mu\nu}}
  & = & \left( -\beta \gt^{\mu\nu}\right) \left( -{1\over 2}
  \sqrt{\gt}\Tt_{\mu\nu}\right)\delta(y) \nonumber \\
  & = & -{\beta\over 2} \sqrt{\gt}(\rhot - 3\pt)\delta(y)\ .
\end{eqnarray}
Putting it together gives
\be
  {\partial \widehat{\cal L}_\phi \over \partial \phi} =
  -{\beta\over 2} \sqrt{{\gt \over g}} (\rhot - 3\pt)\delta(y)\ .
\ee
The $\phi$ equation is thus
\be
  \Box\phi = {\beta\over 4\alpha} {e^{2\beta\phi} \over b} 
  (\rhot - 3\pt)\delta(y)\ ,
\ee
where the D'Alembertian is given by (\ref{box}).
The jump equation for $\phi$ is therefore
\be  \phi_0' = {\beta \over 8\alpha} b_0 e^{2\beta\phi_0} 
  (\rhot - 3\pt)\ .
  \label{jump3}
\ee 

Now we plug in the brane equations to (\ref{f1}), yielding
\be
  {\ddot a_0 \over a_0} + {{\dot a_0}^2 \over a_0^2} =
  -{1\over 576 M_5^6} e^{4\beta\phi_0} \left[
  \left(16 + 3{\beta^2\over \alpha}M_5^3\right) \rhot^2
  + \left(48 - 18{\beta^2\over \alpha}M_5^3\right) \rhot\pt
  + 27 {\beta^2\over \alpha}M_5^3 \pt^2\right]
  - {\alpha \over 3M_5^3} {\dot\phi_0}^2\ .
  \label{f2}
\ee
For a pure cosmological constant on the brane, we have
$\rhot = -\pt = \lambda$.  Then (\ref{f2}) becomes
\be
  {\ddot a_0 \over a_0} + {{\dot a_0}^2 \over a_0^2} =
  {1\over 36 M_5^6}e^{4\beta\phi_0} 
  \left(2-3{\beta^2\over \alpha}M_5^3\right)\lambda^2
  - {\alpha \over 3M_5^3} {\dot\phi_0}^2\ .
\ee
We see that this can vanish for any value of $\lambda$, if 
$\dot\phi_0=0$ and we choose 
\be
  {\beta^2\over \alpha}M_5^3 = {2\over 3}\ .
\ee
This is the condition for self-tuning, as derived by
\cite{Arkani-Hamed:2000eg,Kachru:2000}.  (Of course, making 
this choice is a kind of fine-tuning, as we discuss briefly
in the Appendix.)

In this case, our cosmological equation (\ref{f2}) for more
general matter sources becomes
\be
  {\ddot a_0 \over a_0} + {{\dot a_0}^2 \over a_0^2} =
  - {1\over 32M_5^6} e^{4\beta\phi_0} (\rhot + \pt)^2 
  - {1\over 2}\beta^2 {\dot\phi_0}^2\ .
  \label{f3}
\ee
If the brane tension is $\lambda$,
we can decompose the energy density and pressure into
tension plus dynamical energy-momentum ({\it e.g.} from matter
and radiation on the brane): $\rhot = \lambda + \rhodyn$,
$\pt = -\lambda + \pdyn$.  From (\ref{f3}), however, the
tension simply cancels out, and we are left with an identical
equation relating for the dynamical density/pressure as we
have for the total density/pressure.  As this equation is
quadratic rather than linear in the energy density, it will
not yield anything close to
conventional Friedmann behavior, and can't be taken
seriously as a description of the real world.  (We note that
matter on the brane does not respond directly to the scale factor
$a$, but to the conformally-transformed scale factor; however,
for constant $\phi$ the functional dependence on the energy density will
still be quadratic.)

%%%%%%%%%%%%%%%%%%%%%%%%%%%%%%%%%%%%%%%%%%%%%%%%%%%%%%%%%%%%%%%%%%%%%%%%%
\subsection{Volume-element coupling}
\label{volume}

In this section we consider a model in which $\phi$ enters only
in an overall factor multiplying the Lagrange density, 
not in every appearance of the metric; in the notation of
(\ref{action4}) we have
\be
  \widehat{\cal L}_4 = e^{2\beta\phi} f(g^\4_{\mu\nu}, \psi_i)\ .
\ee
This is equivalent to modifying the four-dimensional
spacetime volume element:
\be
  d^4x\, \sqrt{g^\4} \rightarrow 
  d^4x\, \sqrt{|e^{\beta\phi}g^\4_{\mu\nu}|} =
  d^4x\, e^{2\beta\phi}\sqrt{g^\4} \ ,
\ee
while otherwise coupling to the induced metric $g^\4_{\mu\nu}$.
Unlike the example of conformal coupling, this form of the action 
can be disrupted by quantum
corrections; on the other hand, we shall see that the volume-coupling
ansatz leads to a somewhat more acceptable cosmological model, and
is worth exploring for that reason.

Since $\phi$ now couples non-universally to matter, it is most
sensible to think of $g_{\mu\nu}$ as the metric to which matter
responds, and the coupling to $e^{\beta\phi}$ as an interaction.
The energy-momentum tensor therefore takes the conventional form
\be
  T^\mu{}_\nu = -2 {1\over \sqrt{g^\4}} g^{\4\mu\lambda}
  {\partial \lag_4 \over \partial g^{\4 \lambda\nu}}
  = {\rm diag}(-\rho,p,p,p).  
\ee
This is related to $T^\4_{ab}$, the brane contribution to the
energy-momentum tensor appearing in the five-dimensional Einstein equations,
by
\be
  T^\4_{ab} = -2 {1\over \sqrt{g^\5}}
  {\partial \lag_4 \over \partial g^{\5 ab}}
  = \sqrt{g^\4 \over g^\5} T_{\mu\nu}\delta^\mu_a\delta^\nu_b\ .
\ee
The specific components of $T^\4_{ab}$ are thus:
\begin{eqnarray}
  T^{(4)}_{00} &=& {n^2 \over b} \rho \delta(y) \\
  T^{(4)}_{ij} &=& {a^2 \over b}  p\delta(y)\delta_{ij}\ .
\end{eqnarray}
These differ from (\ref{t400})-(\ref{t4ij}) of the previous section
by the replacement $(e^{2\beta\phi}\rhot, e^{2\beta\phi}\pt)
\rightarrow (\rho, p)$.
As a consequence, the jump conditions for the metric coefficients
are
\bea
  {a_0' \over a_0} &=& -{1\over 6M_5^3} b_0 \rho\ ,\\
  {n_0' \over n_0} &=& {1\over 6M_5^3} b_0
  (2\rho + 3p)\ .
  \label{jumps}
\eea

To express the jump condition for $\phi$ in terms of $\rho$
and $p$, we need to calculate $\partial\widehat{\cal L}_4/\partial\phi$.
We can do this by considering our four-dimensional action to be that
of a perfect fluid, corresponding to
\be
  \widehat{\cal L}_4 = e^{2\beta\phi}f(g^\4_{\mu\nu}, \psi_i) = 
  p(\epsilon, s) - {n \over 2\epsilon}\left(g^\4_{\mu\nu}
  \Omega^\mu \Omega^\nu + \epsilon^2\right)\ 
  \label{perfect}
\ee
(see \cite{Mennim:2000wv,ss} for a discussion). In
(\ref{perfect}), the dynamical degrees of freedom with respect to
which we vary the action to obtain equations of motion include
the entropy $s$, the enthalpy $\epsilon$, 
a Lagrange multiplier $n$, the three Clebsch potentials
$\chi$, $\alpha$, $\beta$, and the thermasy $\theta$; these last four
scalars define the four-vector $\Omega^\mu$ via
\be
  \Omega_\mu = \partial_\mu\chi + \alpha\partial_\mu\beta
  +\theta \partial_\mu s\ .
\ee
Note that varying with respect to $n$ gives the constraint
$\Omega_\mu \Omega^\mu + \epsilon^2 = 0$; hence, on-shell the Lagrange
density is simply equal to the pressure $p$.  
Therefore, from (\ref{phi2}) the equation of motion for $\phi$ is
\bea
  \Box\phi &=& -{1\over 2\alpha} {1\over b}
  {\partial \widehat{\cal L}_4 \over \partial \phi} \delta(y)\\
  &=& -{\beta \over \alpha b}
  p \delta(y)\ .
\eea
This corresponds to a jump condition describing the behavior
of $\phi$ in the vicinity of the brane,
\be
  \phi_0' = -{\beta \over 2\alpha} b_0 p\ .
  \label{phijump2}
\ee

Now we plug in these jump conditions to (\ref{f1}), yielding
\be
  {\ddot a_0 \over a_0} + {{\dot a_0}^2 \over a_0^2} =
  -{1\over 36 M_5^6} \left(\rho^2 + 3\rho p
  + 3 {\beta^2\over \alpha}M_5^3 p^2\right)
  - {\alpha \over 3M_5^3} {\dot\phi_0}^2\ .
  \label{f4}
\ee
We see that the self-tuning condition is satisfied again only for
\be
  {\beta^2\over \alpha}M_5^3 = {2\over 3}\ ,
  \label{tune2}
\ee
in which case our cosmological equation (\ref{f4}) for more
general matter sources becomes
\be
  {\ddot a_0 \over a_0} + {{\dot a_0}^2 \over a_0^2} =
  - {1\over 36M_5^6} (\rho + p)(\rho + 2p)
  - {1\over 2}\beta^2 {\dot\phi_0}^2\ .
  \label{f5}
\ee

The contribution of the brane tension $\lambda$ to the Lagrangian
enters in the combination $e^{2\beta\phi_0}\lambda$.  We 
therefore decompose the
energy density and pressure into tension plus dynamical sources
in the following manner:
\bea
  \rho &=& e^{2\beta\phi_0}\lambda + \rhodyn\ ,\\
  p &=& - e^{2\beta\phi_0}\lambda + \pdyn\ .
\eea
Then, unlike the cosmological equation (\ref{f3}) obtained for 
conformal coupling, in the case of volume-element coupling we
obtain terms on the right-hand side which are linear in 
the dynamical energy density and pressure:
\be
  {\ddot a_0 \over a_0} + {{\dot a_0}^2 \over a_0^2} =
  {1\over 36M_5^6}  \left[e^{2\beta\phi_0}\lambda(\rhodyn + \pdyn)
  - (\rhodyn^2 + 3 \rhodyn \pdyn + 2 \pdyn^2)\right]
  - {1\over 2}\beta^2 {\dot\phi_0}^2\ .
  \label{f6}
\ee
An equivalent equation was derived by Mennim and Battye 
\cite{Mennim:2000wv}.

An equation of this type stands a chance of describing the real
world.  The terms quadratic in energy/momentum can presumably
be neglected at late times; if we assume that the $\dot\phi_0^2$ term 
is negligible we are left with a relation which has some resemblance
to the hoped-for equation (\ref{wish}).  With this in mind, we
turn now to comparison with observation.

%%%%%%%%%%%%%%%%%%%%%%%%%%%%%%%%%%%%%%%%%%%%%%%%%%%%%%%%%%%%%%%%%%%%%%
\section{Discussion}
\label{discussion}

Let us consider whether, under favorable circumstances, (\ref{f6})
could be consistent with what we know about the universe.

A time-dependent bulk scalar $\phi$ would generally lead to 
observable time-dependence in the four-dimensional Newton's
constant, and therefore must be very small.  Although the models
under consideration do not include any mechanism for stabilizing
$\phi$, we will proceed optimistically and 
imagine that the bulk scalar may be approximated as independent
of time,
\be
  \phi(y,t) = \bar{\phi}(y) \ .
\ee
We imagine further that the $y$-dependence of the metric
coefficients may be factored out, and the coefficient $b^2$ of
$dy^2$ set to unity by an appropriate rescaling (which may
always be done if $b^2$ is independent of $t$):
\be
  ds^2 = \omega(y)[-dt^2 + a_0^2(t)d{\vec x}^2] + dy^2\ .
\ee
Referring back to the jump conditions (\ref{jumps}) and
(\ref{phijump2}), such
ans\"atze do not seem unreasonable; in (\ref{jumps})  and
(\ref{phijump2}) the 
symbols $\rho$ and $p$ refer to the entire energy and pressure,
which will be dominated by the brane tension term, which in 
turn remains constant.  Thus, it is reasonable to approximate
the bulk solutions for $\bar\phi$ and $\omega$ by
on their flat-space values as derived in 
\cite{Arkani-Hamed:2000eg,Kachru:2000},
\be
  \bar{\phi}(y) = \phi_0 - {1\over 2\beta}\ln[\omega(y)]
\ee
and
\be
  \omega(y) = \sqrt{1- y/y_c}
  \ ,
\ee
where $y_c$ represents the location of the boundary opposite our brane, 
corresponding to a singularity,
\be
  y_c = {3M_5^3 \over 2\lambda}e^{-2\beta\phi_0}\ .
\ee

In order to compare the modified Friedmann relation
(\ref{f6}) to observations, it is necessary to
express the four-dimensional Planck mass $M_4 = 1/\sqrt{8\pi G_4}$
in terms of five-dimensional quantities.  $M_4$ is defined by
an integral over the extra dimension,
\be
  M_4^2 = M_5^3 \int_0^{y_c} dy\, \omega(y) = {M_5^6 \over \lambda}
  e^{-2\beta\phi_0}\ .
\ee
(In fact, we are again ignoring a subtlety:  due to the presence of
the scalar field, general relativity is not exactly recovered in 
four dimensions, and the definition of $G$ has to be specified more
carefully.  It is possible that solar-system tests of gravity would
rule out these simple models more definitively than the cosmological
scenario considered here.)

For simplicity we drop the subscript ``dyn'' from
$\rho$ and $p$ for the remainder of this section.
In terms of the Hubble parameter 
$H= \dot a/a$, we have $\ddot a/a + (\dot a/a)^2 =
\dot H + 2H^2$.  Then setting $\dot\phi_0=0$ and ignoring 
quadratic terms in $\rho, p$, our cosmological equation 
(\ref{f6}) becomes
\be
  \dot H + 2H^2 = {2\pi G\over 9}(\rho +p)\ .
  \label{f7}
\ee
Unlike the ordinary Friedmann equation, which is a constraint relating
the value of the Hubble parameter to the energy density, this is
a differential equation for $H$, which will involve an additional
integration constant in its solution.  In fact we can integrate (\ref{f7})
explicitly to obtain
\be
  H^2 = {4\pi G\over 9} a^{-4} \int a^3 (\rho + p)da\ .
  \label{f8}
\ee
Consider a universe dominated by a combination of matter
($\rho_\M\propto a^{-3}$, $p_\M = 0$) and radiation 
($\rho_\R\propto a^{-4}$, $p_\R = \rho_\R/3$).  The Hubble parameter
then obeys
\be
  H^2 = {4\pi G\over 9}\rho_\M + {16\pi G\over 27}\rho_\R
  \ln\left(a\over a_*\right)\ ,
  \label{f9}
\ee
where $a_*$ is the integration constant alluded to above.

It is this relation (\ref{f9}) which is to be compared to the conventional
Friedmann law $H^2 = (8\pi G/3)\rho$.  In a matter-dominated
era, the functional dependence of $H$ on $a$ is conventional,
leading to the familiar $a\propto t^{2/3}$ behavior.  The coefficient
of $\rho$, however, is different, suggesting a possible empirical
test of the model: in this theory, the Hubble parameter during matter
domination will be smaller by a factor of $1/\sqrt{6} \approx 0.41$ than
the conventional expectation (at fixed $\rho$).
Unfortunately, we do not have very precise
empirical information about the expansion rate during the matter-dominated
era.   Given that our current universe is apparently dominated by a
smooth component causing it to accelerate, we cannot directly constrain
the coefficient of $\rho_\M$ by contemporary observations.  Since
recombination occurred during the matter-dominated era, it would be
possible in principle to constrain this coefficient via observations
of the CMB anisotropy power spectrum; it is first necessary, however,
to reliably calculate what that power spectrum should be, taking into
account possible long-range deviations from general relativity due to
the extra dimension.  Since we have not undertaken this task, we are
unable to say whether $H^2 = (4\pi G/9)\rho_\M$ is consistent with
the real universe.

In the radiation-dominated era, quantitative constraints on the
behavior of the Hubble parameter may be derived from Big-Bang
Nucleosynthesis (BBN) \cite{bbn}.  Expanding the logarithm in 
(\ref{f9}), we have
\be
  H^2_\R = {16\pi G\over 27}\rho_\R \ln{a} + C a^{-4}\ ,
\ee
where $C$ is an integration constant.  This constant
is not merely a nuisance that can be set to zero, but rather
reflects the normalization of $a$; if we set $a=1$ today, the
first term is always negative in the past, which is clearly
unworkable.  On the other hand, we can choose $C$ such that the
second term is dominant --- and with the correct magnitude to
be compatible with observation --- during BBN.  This seems like
an unlikely bit of fine-tuning, although it cannot be rigorously
excluded.  Therefore we see no way to definitively state that 
the self-tuning cosmologies we have considered are incompatible
with observation, although it would require a mysterious 
coincidence to predict the correct light-element abundances.
Since these models were invented to solve the fine-tuning problem
associated with the cosmological constant, this must be considered
as a strike against them.

Nevertheless, it is important to keep in mind that we have only
dealt with specific toy models of self-tuning, which perhaps it
is too optimistic to expect would lead to completely realistic
cosmologies.  The general idea that the cosmological constant 
problem may be solved not by making the vacuum energy small, but
by making the metric insensitive to its value, is an interesting
one, and it seems worth the effort to attempt to construct
self-tuning models with more acceptable cosmological behavior.

%%%%%%%%%%%%%%%%%%%%%%%%%%%%%%%%%%%%%%%%%%%%%%%%%%%%%%%%%%%%%%%%%%%%%%

\section*{Acknowledgments}

We would like to thank Pierre Binetruy, Gregory Gabadadze,
Steve Giddings, Monica Guica, Shamit Kachru,
Nemanja Kaloper, Finn Larsen, Joe Lykken, Ricardo
Rattazzi, Eva Silverstein, and David Wands for
useful conversations.  This work was supported in part by the
U.S. Dept. of Energy, the Alfred P. Sloan Foundation, and the David
and Lucile Packard Foundation.

%%%%%%%%%%%%%%%%%%%%%%%%%%%%%%%%%%%%%%%%%%%%%%%%%%%%%%%%%%%%%%%%%%%%%%

\section*{Appendix: Two observations}

In this Appendix we very briefly mention two issues of somewhat related
interest to this work: first, the possibility of inflationary behavior
in self-tuning cosmologies, and second, the ability to tune away
other equations of state.

An obvious question which arises in any model which would make
the metric insensitive to vacuum energy is, how are we to explain
the apparent nonzero value of the cosmological constant today
\cite{carroll,sne}, or
implement an inflationary scenario in the early universe?  Of course
if a realistic self-tuning theory is found, it is
conceivable that the currently observed vacuum energy is simply
a reflection of an imperfectly tuned universe, once all aspects
of the theory are taken into account.  On the other hand, it is
interesting to note that there is no difficulty in obtaining
``accelerating'' solutions in the presence of slowly-rolling
scalar fields.  Consider a minimally coupled four-dimensional
scalar $\Phi$, with potential $V(\Phi)$.  The energy density and
pressure are given by
\be
  \rho_\Phi = {1\over 2}\dot\Phi^2 + V(\Phi) \ ,\qquad
  p_\Phi = {1\over 2}\dot\Phi^2 - V(\Phi) \ ,
\ee
or
\be
  \rho_\Phi + p_\Phi = \dot\Phi^2\ .
\ee
For a slowly-rolling scalar with $\dot\Phi \approx$~const,
(\ref{f8}) then implies $H\approx$~const, just as in conventional
theory with a nonzero vacuum energy.  Thus there is no obstacle
in principle to obtaining accelerated expansion either today or
in the early universe.  Of course the usual tuning problems
associated with getting the correct nonzero value of the apparent vacuum
energy are as severe in such a hypothetical model as they are in
conventional quintessence and inflation scenarios.

Another interesting issue is the possibility of choosing
parameters which work to tune away the effects of a more general
energy component with equation of state $p=w\rho$, where $w$
is not necessarily $-1$.  We know of no compelling reason why
this should happen, but the exercise illustrates the extent to
which there really is some tuning going on in our choice of
parameters.

Vacuum energy was tuned away by choosing our parameters $\alpha$
and $\beta$ to be related by (\ref{tune2}) (taking the case of
volume-element coupling for definiteness).  Let us imagine that we
instead take
\be
  {\beta^2\over \alpha}M_5^3 = {2\over 3}x ,
  \label{tune3}
\ee
where $x$ is a parameter to be chosen.  The cosmological equation
(\ref{f4}) becomes (setting $\dot\phi_0 = 0$ for simplicity),
\be
  {\ddot a_0 \over a_0} + {{\dot a_0}^2 \over a_0^2} =
  -{1\over 32 M_5^6} \left(\rho^2 + 3\rho p
  + 2x p^2\right)\ .
  \label{f10}
\ee
For a universe dominated by a component with $p=w\rho$, the
right hand side will automatically vanish if we choose
\be
  x = -{1 + 3w \over 2w^2}\ .
\ee
The choice $x=1$ tunes away vacuum energy ($w=-1$) as well as
an exotic component with $p=-\rho/2$.  With the exception of matter
($w=0$), the expansion can be made insensitive to any specific
equation of state by an appropriate choice of $x$.  Once again,
we have no reason to suggest that such a possibility will occur
(nor see any need for it); however, this phenomenon serves as
an illustration that the specific choice (\ref{tune2}) represents
a kind of fine-tuning in its own right.

\end{document}